\documentclass[aps,twocolumn,superscriptaddress]{revtex4-1}%
  
\usepackage{url}
\usepackage{microtype}
\usepackage[bookmarks]{hyperref}
\hypersetup{colorlinks=true}

\usepackage{amsmath,amsthm,bbm,xfrac}
\usepackage{braket}
\usepackage{bibentry}

\usepackage{booktabs}
\usepackage[]{graphicx}
\usepackage[usenames,dvipsnames,svgnames,table]{xcolor}

\usepackage[dvipsnames]{xcolor}
\usepackage{mdframed}
\usepackage[colorinlistoftodos,textsize=tiny,backgroundcolor=BlueGreen,linecolor=Blue]{todonotes}

\definecolor{dblue}{rgb}{.2,0.2,.8}

\newcommand{\heading}[1]{ \vspace{0.25truecm}\textbf{#1.} }

\newcommand\iii{\text{i}}
\newcommand\tr{\text{Tr}}

\newcommand{\brackets}[3]{\langle #1 | #2 | #3 \rangle}
\DeclareMathOperator{\mspan}{span}

\graphicspath{{pics/}}

\usepackage{mathtools}
\usepackage{verbatim}
\usepackage{enumerate}
\usepackage{graphicx}
\usepackage{hyperref}
\hypersetup{
    colorlinks,
    linkcolor={red!50!black},
    citecolor={blue!50!black},
    urlcolor={blue!80!black}
}
\usepackage{bbm}
\usepackage{booktabs}

\usepackage[caption=false]{subfig}

\begin{document}
\title{Complex Networks from Classical to Quantum}

 \author{Jacob Biamonte}\email{jacob.biamonte@qubit.org}\homepage{DeepQuantum.AI}  
\affiliation{Deep Quantum Labs \\ Skolkovo Institute of Science and Technology, Skoltech Building 3, Moscow Russia 143026}
\author{Mauro Faccin}\email{mauro.faccin@uclouvain.be} 
\affiliation{ICTEAM, Universit{\'e} Catholique de Louvain, Euler Building 4, Avenue Lemaitre, B-1348 Louvain-la-Neuve, Belgium}

\author{Manlio De Domenico}\email{mdedomenico@fbk.eu} 
\affiliation{Fondazione Bruno Kessler, Via Sommarive 18, 38123 Povo (TN), Italy}

\maketitle 

{\bf
Recent progress in applying complex network theory to problems in quantum information has resulted in a beneficial crossover.
Complex network methods have successfully been applied to transport and entanglement models while information physics is setting the stage for a theory of complex systems with quantum information-inspired methods.
Novel quantum induced effects have been predicted in random graphs---where edges represent entangled links---and quantum computer algorithms have been proposed to offer enhancement for several network problems.
Here we review the results at the cutting edge, pinpointing the similarities and the differences found at the intersection of these two fields.
}


\vspace{0.25truecm}

Quantum mechanics has long been predicted to help solve computational problems in physics \cite{2014arXiv1405.2831J}, chemistry \cite{2010NatCh...2..106L}, and machine learning \cite{2016arXiv161109347B} and to offer quantum security enhancement in communications~\cite{komar2014quantum}, including a quantum secure Internet~\cite{kimble2008quantum}.   Rapid experimental progress has pushed quantum computing and communication devices into truly data-intensive domains, where even the classical network describing a quantum system can exhibit complex features, giving rise to what appears as a paradigm shift needed to face a fundamental type of complexity \cite{acin2007entanglement, faccin2013degree, QuantumPageRank, garnerone2012pagerank, paparo2014google, sanchez2012quantum,lu2014chiral, faccin2014community}.  Methods originating in complex networks---traditionally based on statistical mechanics---are now being generalized to the quantum domain in order to address these new quantum complexity challenges.  

Building on several fundamental discoveries~\cite{WS98,barabasi1999emergence}, complex network theory has demonstrated that many (non-quantum) systems exhibit similarities in their complex features~\cite{WS98,barabasi1999emergence,boccaletti2006complex,kivela2014multilayer,dedomenico2016physics}, in the organization of their structure and dynamics~\cite{guimera2005functional,palla2005uncovering,song2005self,colizza2006detecting,boguna2009navigability,vespignani2012modelling}, the controllability of their constituents~\cite{liu2011controllability} and their resilience to structural and dynamical perturbations~\cite{albert2000error,callaway2000network,buldyrev2010catastrophic,gao2012networks,radicchi2013abrupt,dedomenico2014navigability}. Certain quantum systems have been shown to indeed exhibit complex features related to classical systems, as well as novel mechanisms and principles that interrelate complex features in quantum systems \cite{acin2007entanglement, faccin2013degree, QuantumPageRank, garnerone2012pagerank, paparo2014google, sanchez2012quantum,lu2014chiral, dedomenico2016spectral}.  

Two types of quantum networks have been of primary focus in the series of pioneering results we review.  The first consists of quantum systems whose connections are represented by entangled states~\cite{acin2007entanglement, cuquet2009entanglement, Perseguers2010}.  These quantum networks are used in secure quantum communication systems.  The second area of focus consists of networks of quantum systems, such as atoms or superconducting quantum electronics, whose connections are physical \cite{cirac1997quantum,chaneliere2005storage,wilk2007single,politi2008silica,ritter2012elementary,aspuru2012photonic}.  Such systems are used to develop quantum-enhanced algorithms or quantum information transport systems, both modeled by quantum walks on complex networks.  At a fundamental level, the two types of quantum networks are described by quantum information theory, allowing one to extend the spectrum of network descriptors---such as ranking indicators, similarity and correlation measures---inside the quantum domain. 

\begin{figure*}[!t]
  \begin{mdframed}[frametitle=Box 1 \rule{2pt}{7pt} Cross-pollination between the fields of complex networks and quantum information science.,
    outerlinewidth=2,
    frametitlebackgroundcolor=yellow!20]
  
    In recent years, seminal work has been carried out at the intersection of quantum information and computation and complex network theory.
    We attempt to catalog the scope of this work  in Fig.~\ref{fig:qnet}.

\includegraphics[width=\textwidth]{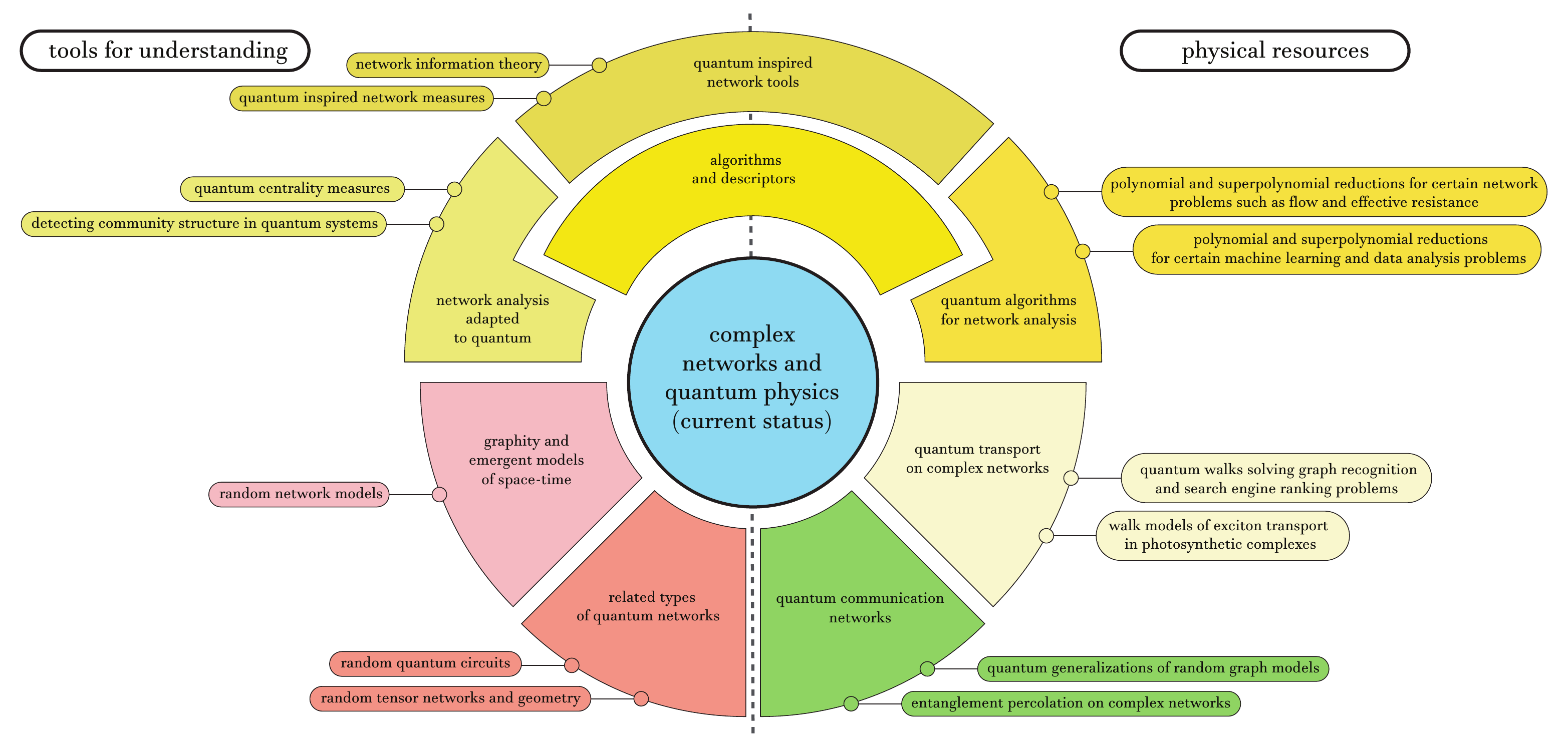}
\caption{Each of the shaded regions represent published findings that map out the field from theoretical, experimental and computational perspectives. The top area classifies the analytical tools inspired by quantum information for classical network analysis and vice versa---both covered in this review---as well as the quantum algorithms developed to address specific problems in network science. The classification in the bottom area includes quantum networks based on entangled states and on physical connections---also covered in this review---as well as quantum network models of space-time, random quantum circuits, random tensor networks and geometry.}
\label{fig:qnet}
\end{mdframed}
\end{figure*}

Interestingly, the same tools can then be appropriately modified to apply to traditional complex networks, suggesting the existence of a framework---network information theory---suitable for application to both classical and quantum networked systems~\cite{dedomenico2015structural,dedomenico2016spectral,schieber2016quantification}. This bidirectional cross-over is carving out a coherent path forward built fundamentally on the intersection of these two fields (see Fig.~\ref{fig:qnet}). Several quantum effects are still outside of the predictive range of applicability of complex network theory.  Future work should build on recent breakthroughs and head towards a new theory of complex networks which augments the current statistical mechanics approach to complex networks, with a theory built fundamentally on quantum mechanics.  Such a unified path forward appears to be through the language of information theory.  

Here, we make an effort to review some of the crucial steps towards the creation of a network theory based fundamentally on quantum effects. Therefore, we do not cover several topics that, nevertheless, deserve to be mentioned as part of the field. These include, in no particular order, quantum gravity theories based on complex networks~\cite{ambjorn2004emergence,levin2005string,konopka2008quantum,rovelli2010geometry}, synchronization in and on quantum networks~\cite{vinokur2008superinsulator}, quantum random circuits~\cite{emerson2003pseudo,brown2010convergence}, classical spin models and quantum statistics successfully used in complex network theory~\cite{bianconi2001bose,reichardt2004detecting,garlaschelli2009bosefermi} (see \cite{dorogovtsev2008critical} for a thorough review).

\section*{Networks in quantum physics vs complexity} 

Network and graph theory fundamentally arises in nearly all aspects of quantum information and computation.   As is the case with traditional network science, not all networks exhibit what is considered as `complexity'.  Here we will recall briefly the basic definition of a network and mention several areas where network theory arises in quantum computation and contrast this with the concept of a complex network.  
 
A network is an abstract representation of relationships (encoded by edges) between units (encoded by nodes) of a complex system. Edges can be directed, i.e.~they can represent information incoming to or outgoing from a node and, in general, they can be weighted by real numbers. The number of incoming, outgoing and total edges is known as incoming, outgoing and total degree of a node, respectively. The sum of the corresponding weights defines the incoming, outgoing and total strength of that node, respectively. Networks are often characterized by how node degree and strength are distributed and correlated. Systems modeled by uncorrelated networks with homogeneous degree distribution are known as Erdos-Renyi networks, whereas systems with power-law degree distribution are known as scale-free networks. We refer to \cite{Newman03thestructure,boccaletti2006complex} for reviews of network concepts and models.

The use of various aspects of graph and network theory can be found in all aspects of quantum theory, yet not all networks are complex.  The commonly considered networks include (i) Quantum spins arranged on graphs; (ii) quantum random walks on graphs; (iii) Quantum circuits/networks; (iv) Superconducting quantum (electrical) circuits; (v) Tensor network states; (vi) Quantum graph states, etc.  Although the idea of a complex network is not defined in a strict sense, the definition is typically that of a network which exhibits an emergent property, such as a non-trivial distribution in node degree.  This is in contrast to graph theory, which applies graph theory or tensor network reasoning to deduce and determine properties of quantum systems.  Here we will focus on topics in quantum systems which are known to be connected with the same sort of complexity considered in complex networks.  

\section*{Quantum networks based on entangled states}

To define quantum networks based on entangled states, let us start from the state of each $i^{\text th}$ qubit, written without loss of generality, as
\begin{eqnarray}
\ket{\psi_i} = \cos(\alpha_{i})\ket{0} + e^{-i \theta_i}\sin(\alpha_{I})\ket{1},
\end{eqnarray}
with $\ket{0}$ and $\ket{1}$ the preferred or `computational' basis. The qubit is in a pure, coherent superposition of the two basis states and any measurement in this same basis will cause the state to collapse onto $\ket{0}$ or $\ket{1}$, with probability $\cos^2(\alpha_{i})$ and $\sin^2(\alpha_{I})$, respectively. Let us consider a quantum system with two qubits,~i.e. $i=1$ and 2. The basis of this system is given by the so called, tensor product, of the two basis states: $\ket{00}$, $\ket{01}$, $\ket{10}$ and $\ket{11}$. If the two qubits are not entangled, i.e.~their states are independent from each other, then the state of the overall system can be written as e.g.~$\ket{\psi_{12}}=\ket{\psi_1}\otimes\ket{\psi_2}$, whereas this is not possible if the two qubits are entangled. A generalization of this description to the case of mixed states is obtained in terms of the non-negative density matrix $\rho$; a unit trace Hermitian operator representing the state of the system as an ensemble of (unknown) pure states.

Instead of distributing entanglement on regular graphs, such as uniform lattices typically studied in condensed matter physics, it has been shown that it is possible to tune the amount of entanglement between two nodes in such a way that it equals the probability to have a link in (classical) Erdos-Renyi graphs~\cite{acin2007entanglement}. Such random graphs can be defined by the family of networks $G(N,p)$, where $N$ is the number of nodes and $p$ the probability to find a link between any two nodes.
The probability scales with the size of the network following a power law $p\propto N^{-z}$, with $z\geq 0$.
In classical network theory, there exists a critical value for the probability $p_c(N)$ for which, if $p>p_c(N)$ a given subgraph of $n$ nodes and $l$ links has higher probability to be observed. The classical result is that this critical probability scales with $N$ as $p_c(N)\propto N^{-n/l}$.

\begin{figure}[ht]
\centering
\includegraphics[width=\columnwidth]{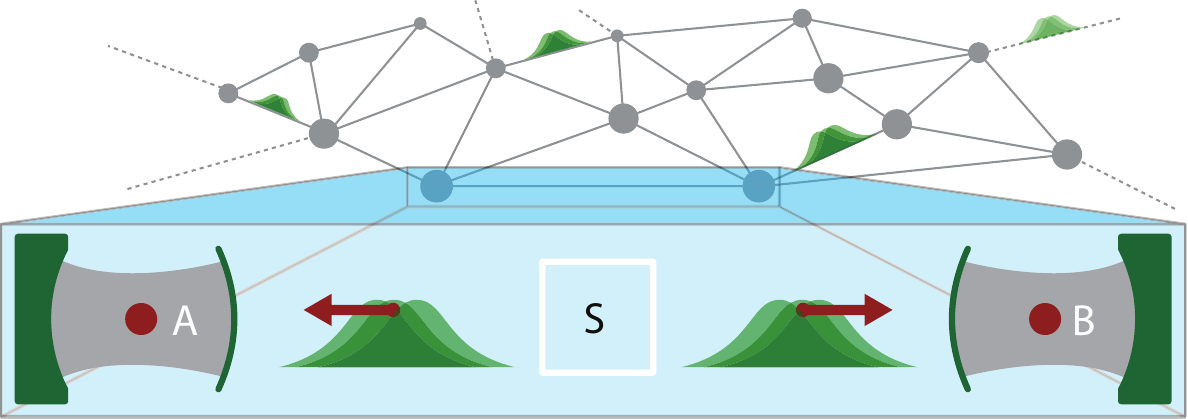}
\caption{{\bf Entanglement connecting distant network nodes}. A pair of distant
cavities in a quantum communication network are driven by a shared squeezed light source $S$.  In the systems steady state the two atoms $A$ and $B$ entangle, forming a network edge~\cite{Perseguers2010}. 
}\label{fig:entanglement}
\end{figure}

Acin, Cirac, and Lewenstein~\cite{acin2007entanglement} formulated an elegant extension of this picture to the quantum realm by replacing each link with an entangled pair of particles, where the probability $p_{i,j}=p$ that the link exists between nodes $i$ and $j$ is substituted by a quantum state $\rho_{i,j}:=\rho$ of two qubits, one at each node  (see Fig.~\ref{fig:entanglement}).
One can build a quantum network where each node consists of $N-1$ qubits which are entangled, in pairs, with qubits of other nodes.
However, in this case, although the connections are identical and pure they encode non-maximally entangled pairs. 
For pure states of qubits the density matrix is $\rho=\ket{\phi}\bra{\phi}$, with
\begin{eqnarray}
\ket{\phi} = \frac 1{\sqrt{2}} \left( \sqrt{2-p} \ket{00} + \sqrt{p} \ket{11} \right).
\end{eqnarray}
Here, $0\leq p\leq 1$ quantifies the entanglement of links and the state of the overall quantum random graph can be denoted by $|G(N,p)\rangle$. If each link, i.e.~each entangled pair, attempts to convert its state to the maximally entangled one ($p = 1/2$)
through local operations and classical communication (LOCC), the optimal probability of successful conversion is exactly $p$. It follows that the fraction of existing entangled states converted to maximally entangled ones by LOCC corresponds to the probability of having a link between nodes in the corresponding classical random network~\cite{acin2007entanglement}. By varying the value of the parameter $z$, i.e.~how the critical probability scales with system size, it is possible to control the number and type of subgraphs present in a quantum network of $N$ nodes. This is useful to create special multipartite states, such as the Greenberger-Horne-Zeilinger state which exhibits non-classical correlations~\cite{greenberger1989going}. The striking result is that it is possible to obtain, with probability approaching unity, a quantum state with the topology of any finite subgraph for $N$ approaching infinity and $z=-2$.

This bridge between complex network theory and quantum theory provides a
powerful tool to investigate the critical properties of a quantum system. For
instance, in the case of regular lattices, it has been shown that the
probability $p_\text{opt}$ to establish a perfect quantum channel between the nodes can be mapped to the probability of distributing links among each pair of nodes in the lattice~\cite{acin2007entanglement}, a scenario that can be studied using the well-established bond-percolation theory from statistical physics. This result allows one to calculate the critical probability above which the system will exhibit an infinite connected cluster and, in the case of qubits, it has been shown that the probability of having an entangled path with infinite length---i.e., an infinite sequence of entangled states connecting an infinite number of qubits---is unity.  However, for product states this probability is zero, denoting the existence of a sharp transition between these two scenarios. However, local measurements based on this approach, called classical entanglement percolation (CEP), are not optimal, in general, to generate maximally entangled states: CEP is not even asymptotically optimal for two-dimensional lattices and new quantum protocols based on quantum entanglement percolation have to be used instead~\cite{acin2007entanglement}. A novel critical phenomenon, defining an entanglement phase transition, emerges from this new strategy, where the critical parameter is the degree of entanglement required to be distributed in order to establish a quantum channel with probability that does not decay exponentially with the size of the system, at variance with CEP. This type of enhancement with respect to the classical case has been reported for different network topologies, such as Erdos-Renyi, scale-free and small-world networks~\cite{cuquet2009entanglement}.

Unexpected quantum effects emerging from network effects have been reported. Cardillo et al. show that nodes which store the largest amount of information are the ones with intermediate connectivity and not the hubs, breaking down the usual hierarchical picture of classical networks~\cite{cardillo2013information}. More recently, Carvacho et al. measured the emergence of special quantum correlations, named non-bilocal, correlating distant qubits by means of several intermediate, typically independent, sources, and providing evidence for violation of local causality in a quantum network~\cite{carvacho2017violation}.

The static entangled states providing the network connectivity described here, will be replaced in the next section by dynamical processes on networked quantum systems.  



\section*{Quantum networks based on physical connectivity}

Another wide area where network concepts have found applicability consists of quantum systems physically interconnected, such as atoms or superconducting quantum electronics \cite{cirac1997quantum,chaneliere2005storage,wilk2007single,politi2008silica,ritter2012elementary,aspuru2012photonic}.  These types of systems provide fertile ground where quantum algorithms are tested \cite{ambainis2003quantum, shenvi2003quantum, shenvi2003quantum, rossi2015measuring, 2016arXiv160305423W, 2017arXiv170104392C} and quantum information transport systems are studied \cite{QuantumWalks, Whitfield2010, zimboras2013quantum}.  Typical modeling approaches are based on so called `quantum walks' on complex networks, with recent studies showing that quantum information tasks, typically designed for simple topologies, retain performance in very disordered structures~\cite{2016PhRvL.116j0501C}.  Stochastic (non-quantum) walks are also a central model in complex network theory---see the review \cite{2016arXiv161203281M}. 

Any quantum process can be viewed as a single particle walk on a graph.  Single-particle quantum walks represent a universal model of quantum computation----meaning that any algorithm for a quantum computer can be translated into a quantum walk on a graph---and, additionally, quantum walks have been widely studied in the realm of quantum search on graphs, in both continuous and discrete time via coined walks (see e.g.~\cite{ambainis2003quantum, shenvi2003quantum, shenvi2003quantum, rossi2015measuring, 2016arXiv160305423W}---in particular the graph optimality results~\cite{2016PhRvL.116j0501C}).  The computational advantages of quantum versus stochastic random walk based algorithms has attracted wide interest with typical focus being on general graphs which consequently do not exhibit complex features. However, many works have compared properties of stochastic~\cite{noh2004random, burda2009localization} and quantum random walks \cite{QuantumWalks, Whitfield2010, zimboras2013quantum} on complex networks \cite{2016PhRvE..93b2304M}.  

Network topology has further been shown as a means to direct transport by adding complex numbers---while maintaining Hermiticity---to the networks adjacency matrix in `chiral quantum walks' \cite{zimboras2013quantum, cameron2014universal, 2016t160600992T} (note that chiral walks were realized experimentally in \cite{lu2014chiral}).  Open system walks which mix stochastic and quantum effects in `open' evolutions \cite{Whitfield2010} have aided in the study of quantum effects in biological exciton transport (again, modeled as a quantum walk) and developments in a quantum version of Google's PageRank \cite{QuantumPageRank, garnerone2012pagerank, paparo2014google} has been seen, providing a practical solution to overcome the degeneracy issues affecting the classical version and enhancing node ranking in large networks.  Recently, Faccin et al.~have analytically solved a model which shed light on some key differences between stochastic and quantum walks on complex networks \cite{faccin2013degree}.  These differences push forward a general understanding which can lead to a theory explaining novel complex features in quantum systems.  

Quantum walks on complex networks represent both a practical model of transport \cite{2016PhRvE..93b2304M} as well as an interesting stage of comparison between the quantum and stochastic cases.  As a closed quantum system exhibits fluctuations in the probabilities in time, typically a long time average is considered.  Physically, this is the best approximation one can hope for, provided that there is no knowledge of when the walk started.  In this case, the probability to find a quantum walker in the $i$-th node is given by 
\begin{equation}
  p(i) = \lim_{T\to\infty} \frac 1T \int_0^T \textrm{d}t
  |\braket{i|U_t|0}|^2,
  \label{eq:longtimeprob}
\end{equation}
where $\ket{0}$ is the initial state and $U_t=e^{-\iii Qt}$ 
is the unitary evolution operator defined by the quantum generator $Q$.

Interference between subspaces of different energy vanish in the long time average so we obtain an expression for the probability $ ( P_Q )_i$ in terms of the energy eigen-space projectors $\Pi_j$ of the Hamiltonian $H_Q$, 
\begin{equation}
  ( P_Q )_i = 
     \sum_j \brackets{i}{ \Pi_j \rho (0) \Pi_j}{i} .
     \label{eq:pn}
\end{equation}	
Here $\Pi_j = \sum_k\ket{\phi_j^k}\bra{\phi_j^k}$ projects onto the subspace spanned by the eigenvalues $\ket{\phi_j^k}$ of $H_Q$ corresponding to the same eigenvalue $\lambda_j$. 

\heading{Quantum-enhanced page-ranking}  The non-symmetric adjacency matrix
representing the directed connectivity of the World Wide Web, a.k.a.~the
Google matrix $G$, satisfies the Perron-Frobenius theorem \cite{baezbook} and
hence there is a maximal eigenvalue corresponding to an eigenvector of
positive entries $Gp = p$. The eigenvector $p$ corresponds to the limiting
distribution of occupation probabilities of a random web surfer---it
represents a unique attractor for the dynamics independently of the initial
state. The vector $p$ is known as the Page-Rank.%
\footnote{a dumping or teleportation factor is often included in
the computation in order to assure the Perron-Frobenius theorem
satisfactibility.}
 
Several recent studies embed $G$ into a quantum system and consider quantum versions of Google's Page-Rank \cite{QuantumPageRank,garnerone2012pagerank, paparo2014google}. Garnerone et al.~\cite{Garnerone2012google} relied on an adiabatic quantum algorithm to compute the Page-Rank of a given directed network, whereas Burillo et al.~\cite{sanchez2012quantum} rely on a mixture of unitary and dissipative evolution to define a ranking that converges faster than classical PageRank.

The page-ranking vector $\vec p$ is an eigenvector of $\mathbbm{I}-G$ corresponding to the zero eigenvalue (the lowest). This fact leads to a definition of a Hermitian operator which can play the role of a Hamiltonian, defined as:
\begin{equation}
  h^p = (\mathbbm{I}-G)^\dagger(\mathbbm{I}-G),
\end{equation}
though highly non-local, its ground state represents the target Page-Rank which could be found by adiabatic quantum annealing into the ground state.  Using a quantum computer to accelerate the calculation of various network properties has been considered widely~\cite{ambainis2003quantum, shenvi2003quantum, shenvi2003quantum, rossi2015measuring, 2016arXiv160305423W}.  As Page-Rank relies on finding the vector corresponding to the lowest eigenvalue of the Google matrix, the adiabatic algorithm opens the door up to accelerate network calculations using quantum computers.

\heading{Directing transport by symmetry breaking in chiral walks} Chiral
quantum walks, introduced by Zimbor\'as et al.~in~\cite{zimboras2013quantum}
and realized experimentally in \cite{lu2014chiral}, append complex numbers to
the adjacency matrix (playing the role of the system Hamiltonian) while still
maintaining the Hermitian property~\cite{zimboras2013quantum,lu2014chiral,
cameron2014universal, 2016t160600992T}.  These complex phases in many cases do
not affect transfer probabilities: the theory explaining this finding was developed in~\cite{zimboras2013quantum,lu2014chiral}, without relying on approximations or averaging. The case of open systems has been investigated as well in \cite{zimboras2013quantum}. In the scenarios where the addition of complex phases affects transfer probabilities, the underlying system breaks time-reversal symmetry and, consequently, the probability flow into the quantum system is biased. This fact enables directed state transfer
without requiring a biased (or non-local) distribution in the initial states, or coupling to an
environment.

When the underlying graph is bipartite (e.g.~a graph whose vertices can be divided into two disjoint sets such as a square lattice), time-reversal symmetry in the transport probabilities can not be broken.  Transport suppression is indeed possible however~\cite{zimboras2013quantum}.  Bipartite graphs include trees, linear chains and generally, graphs with only even cycles.  These results point to a subtle interplay between the topology of the underlying graph, giving rise to a new challenge for dynamical control of probability transfer when considering walks on complex networks~\cite{zimboras2013quantum,lu2014chiral, cameron2014universal, 2016t160600992T}.

\heading{Open quantum walks}  The area of open quantum systems \cite{breuer2007theory} studies noise and its effects in quantum systems.  The adiabatic version of Page-Rank~\cite{garnerone2012pagerank} uses a quantum stochastic quantum walk as proposed by~\cite{Whitfield2010} (see also \cite{PhysRevA.91.042108, 2014PhRvB..90l5138M} for studies on open walks).  
 
Quantum stochastic walks are defined by a quantum walk undergoing dissipative dynamics. The latter follows the quantum master equation in the Lindbladian form:
\begin{align}
  \dot\rho &= \mathbbm{L}[\rho]\\ 
           &= -\iii[H,\rho] 
             + \sum_k L_k\rho L_k^\dagger 
             - \frac 12 \left\{ L_k^\dagger L_k,\rho \right\}
  \label{eq:google-dissipation}
\end{align}
where $L_k$ represents a jump operator while $[\cdot,\cdot]$ and $\{\cdot,\cdot\}$ are commutator and anti-commutator respectively. 

The network topology is embedded by choosing $H$ equal to the adjacency matrix of
the symmetrized network and $L_k = L_{ij} = \sqrt{G_{ij}}\ket{i}\bra{j}$.
In this picture, node ranking is defined by an activity vector $\alpha$ computed at the steady state $\rho^\text{ss}$. 

Paparo et al.~\cite{QuantumPageRank,paparo2014google} introduced a Szegedy
type of Markov chain quantization~\cite{szegedy2004quantum} of the random walk.
In order to quantize the Markov chain defined by the Google matrix $G$ of
$N$ nodes, one introduces a Hilbert space 
$\mathcal{H}=\mspan \{\ket{i}_1\ket{j}_2, i,j \in [0,N]\}$ and the
superposition of outgoing edges from node $i$:
\begin{equation}
  \ket{\psi_i} = \ket{i}_1 \otimes \sum_k \sqrt{G_{ki}} \ket{k}_2
\end{equation}
and
\begin{equation}
  \Pi = \sum_k \ket{\psi_k}\bra{\psi_k}.
\end{equation}

Each step of the quantum walk $U$ is defined by a coin flip $2\Pi-\mathbbm{1}$
and a swap operation $S$ which ensures unitarity~\cite{QuantumPageRank}
\begin{equation}
  U = S(2\Pi-\mathbbm{1})
\end{equation}
the swap operator is $S=\sum_{ij} \ket{ij}\bra{ji}$.

In the case of quantum Page-Rank, this is set to the instantaneous probability
$P(i,t)$ of finding the walker at node $i$ at the time-step $t$. 
To obtain a fixed value for the quantum Page-Rank a time average is
calculated as long as with its variance as a measure for quantum fluctuations.

Another approach~\cite{sanchez2012quantum} involves defining a Markov
quantum evolution similar to Eq.~\eqref{eq:google-dissipation}, with a tuning
parameter $\alpha$:
\begin{equation}
  \dot\rho = -\iii(1-\alpha) [H,\rho]
             +\alpha \left[\sum_k L_k\rho L_k^\dagger 
             -\frac 12 \{L_k^\dagger L_k,\rho\}\right]
\end{equation}
where the Hamiltonian $H$ is the symmeterized adjacency matrix and 
$L_k = L_{ij} = \sqrt{G_{ij}}\ket{i}\bra{i}$ represent the jump operators
which consider the directness of the network.
With this definition, for values of $\alpha\in (0,1]$, a stationary state is
guaranteed.
In this case, for $\alpha=0$ we revert to the unitary evolution while for
$\alpha=1$ we revert to the stochastic case. 

The authors~\cite{sanchez2012quantum, QuantumPageRank,paparo2014google} show how this definition of quantum Page-Rank resolves problems
of degeneracy in the classical Page-Rank definition, enhances the
importance of secondary hubs and, for certain values of $\alpha$, the
algorithm exhibits faster convergence.   

\begin{figure}[htb]
  \centering
  \includegraphics[width=0.8\columnwidth]{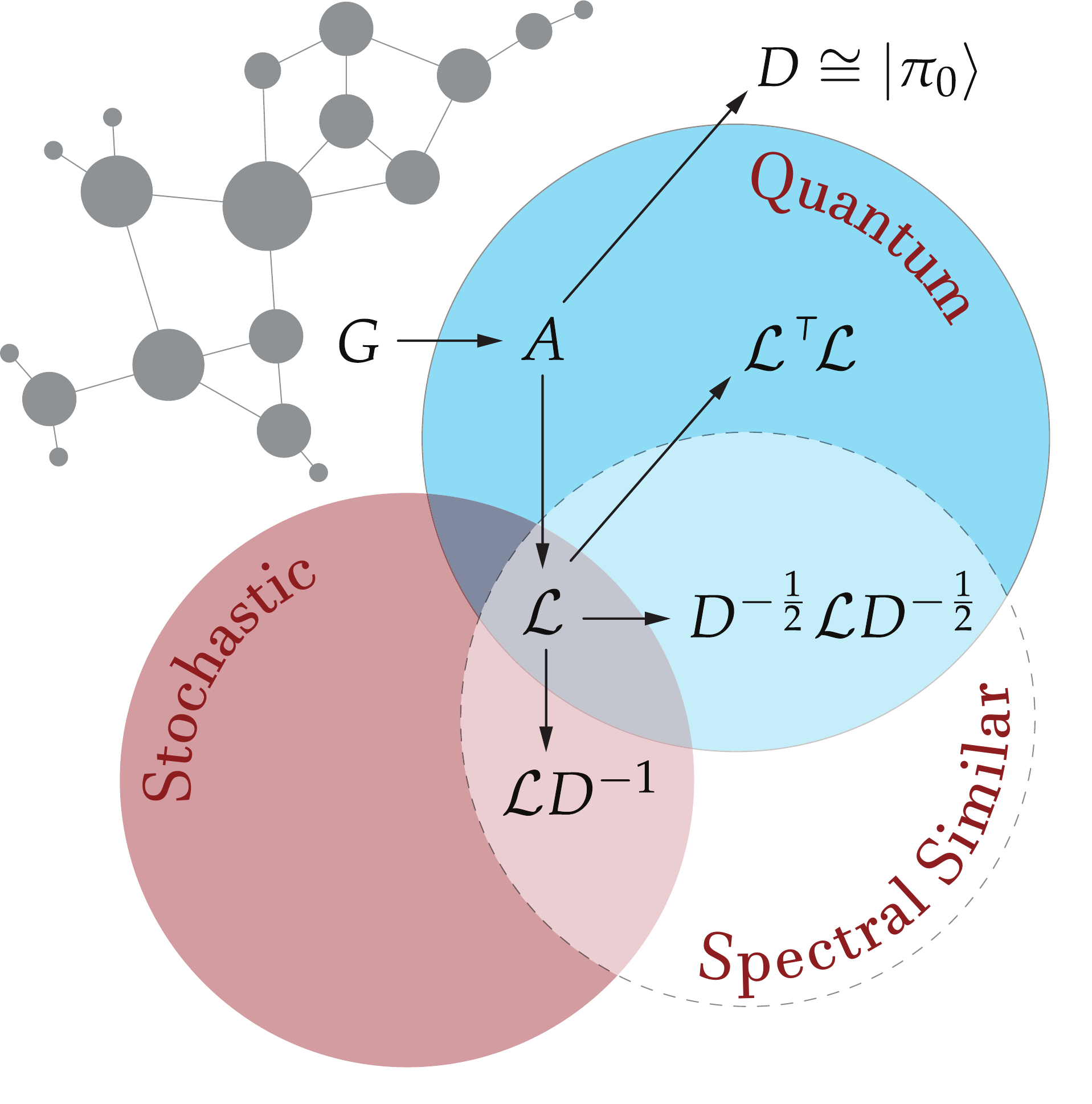}
  \caption{{\bf Known mappings between quantum and stochastic generators.} Here $G=(V,E)$ is a graph with adjacency matrix $A$, $D$ is a diagonal matrix of node degrees.  This yields the graph Laplacian   $\mathcal{L}=D-A$, and hence, the stochastic walk generator $L_{S}=\mathcal{L}D^{-1}$, from this a similarity transform results in $L_{Q}=D^{-\sfrac 12}\mathcal{L}D^{-\sfrac 12}$, which generates a valid quantum walk and exhibits several interesting connections to the classical case.   The mapping $\mathcal{L}\longrightarrow \mathcal{L}^\top\mathcal{L}$ preserves the lowest $0$ energy ground state, opening the door for adiabatic quantum annealing which solves computational problems by evolving a system into its ground state.  }
\label{fig:similarity}
\end{figure}

\section*{Towards unified analysis of network complexity}

The interaction between network science and quantum information science has led to the development of theoretical and computational tools that benefitted from both fields. On the one hand, quantum-inspired tools, such as information entropies and quantum distance measures, have been successfully applied to practical problems concerning classical complex networks~\cite{dedomenico2016spectral}. On the other hand, classical network descriptors have been ported to the quantum realm to gain better insights about the structure and the dynamics of networked quantum systems~\cite{faccin2014community}. The cross-pollination between the two fields---including, among others, quantum statistics for modeling the dynamics of classical networks and their geometry~\cite{javarone2013quantum,bianconi2016network} (see also Ref.~\cite{bianconi2015interdisciplinary} and references therein)---is still ongoing with vibrant future research opportunities. Here we briefly review the advances concerning quantum-inspired entropic measures for networks and network-inspired measures for quantum systems.

\heading{Information entropy of classical networks} Historically, the concept of entropy has been successfully used to quantify the complexity of many systems~\cite{pincus1991approximate,costa2002multiscale}. Recently, the possibility of using quantum entropy and other quantum information theoretical measures has been explored by the community of network scientists. 

For classical complex networks, von Neumann's entropy has been applied over one decade ago~\cite{braunstein2006laplacian}. The combinatorial Laplacian matrix $L$, obtained from the adjacency matrix representing the network, is rescaled by the number of edges in the network. The normalization of the matrix $L$ guarantees that the corresponding eigenvalues are non-negative and sum up to 1---in order to be interpreted as probabilities~\cite{anand2011shannon}---and some other properties which makes the resulting object similar to a quantum density matrix $\rho$. Network entropy is defined according to von Neumann quantum entropy as
\begin{eqnarray}
S(\rho)=-\tr{(\rho \log_{2} \rho)}.
\end{eqnarray}
By exploiting the eigen-decomposition of the Laplacian matrix, it can be shown that this entropy corresponds to the Shannon entropy of the eigenvalue spectrum of $\rho$. This entropy has been generalized to the case of multilayer systems~\cite{dedomenico2013mathematical}, composite networks where units exhibit different types of relationships that are generally modeled as different layers (see~\cite{kivela2014multilayer,boccaletti2014structure,dedomenico2016physics} for a thorough review).

It has been recently shown that the von Neumann entropy calculated from the rescaled Laplacian does not satisfy the sub-additivity property in some circumstances~\cite{dedomenico2016spectral, dedomenico2015structural}. 
This undesirable feature can be addressed by means of a more grounded definition~\cite{dedomenico2016spectral}, whose rationale is to measure the entropy of a network by exploiting how information diffuses through its topology. Information diffusion in this context is governed by the equation
\begin{eqnarray}
\dot{\psi}_{i}(t) = -\sum_{j=1}^{N}L_{ji}\psi_{j}(t),
\end{eqnarray}
with $\psi_{i}(t)$ the amount of information in node $i$ at time $t$. The solution of this diffusion equation is given, in vector notation, by $\boldsymbol{\psi}(t)=\exp(-L t)\boldsymbol{\psi}(0)$, whose normalized propagator is used to define the density matrix as
\begin{eqnarray}
\label{eq:def-rho}
\rho = \frac{e^{-\tau L}}{\tr{(e^{-\tau L})}},
\end{eqnarray}
where time plays the role of a resolution parameter allowing one to probe
entropy at different scales~\cite{dedomenico2016spectral}. A similar approach,
involving a modified Laplacian matrix, has been recently used for revealing
the mesoscale structure of complex directed
networks~\cite{fanuel2016magnetic, fanuel2016visualization}. 

This quantum-inspired framework provides a powerful basis to develop an information theory of complex networks, with direct applications in classical network science, such as system comparison.

\heading{Comparing classical networks}  A known problem in network science is to compare two networks, without relying on a specific subset of indicators. Network information entropy allows one to introduce relative entropies such as the Kullback-Leibler divergence, to compare two networks with density matrices $\rho$ and $\sigma$ respectively:
\begin{eqnarray}
\label{eq:kld-quantum}
\mathcal{D}(\rho||\sigma)=\tr{[\rho(\log_{2}\rho-\log_{2}\sigma)]}.
\end{eqnarray}
By exploiting the well-known classical result that the minimization of Kullback-Leibler divergence between a reference distribution and its parametric model corresponds to the maximization of the likelihood, it has been shown that in a network context this allows one to define the network log-likelihood by
\begin{eqnarray}
\label{eq:ll}
\log_{2}\mathcal{L}(\Theta)=\tr{[\rho\log_{2}\sigma(\Theta)]}.
\end{eqnarray}
The introduction of network likelihood opens the door to a variety of applications in statistical inference and model selection, based on concepts such as the Fisher information matrix, Akaike and Bayesian information criteria, and minimum description length, to cite some of them~\cite{dedomenico2016spectral}.

This new framework has been used to compare networks for several purposes. For instance, in the case of pairs of networks, the graphs are first merged by connecting each node from one network to any other node in the other network. Successively, continuous-time quantum walks are used to explore the composite system and the quantum Jensen-Shannon divergence between the evolution of two walks is calculated. This divergence, that is a measure of (dis)similarity, is shown to be maximum when the two original networks are isomorphic~\cite{rossi2015measuring}.

\begin{figure*}[!t]
\centering
\includegraphics[width=.8\textwidth]{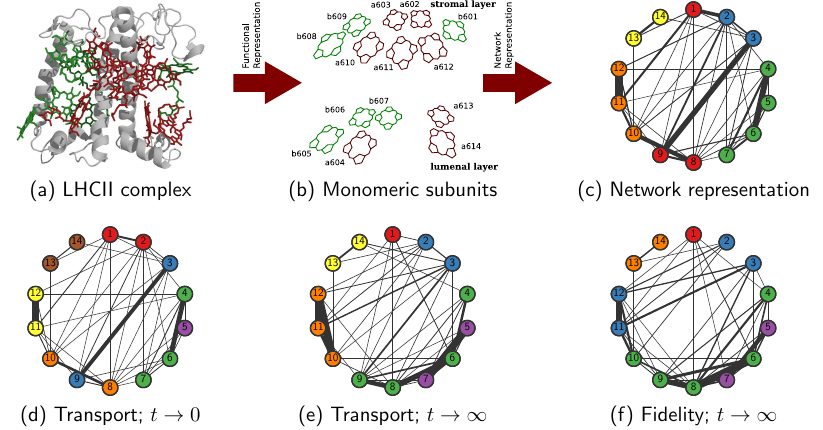}
\caption{%
  \textbf{Community Detection in a light harvesting network (LHCII)}~\cite{faccin2014community}.
  (a) Monomeric subunit of the LHCII complex with pigments Chl-a (red) and Chl-b (green) packed in the protein matrix (gray).
  (b) Schematic representation of Chl-a and Chl-b in the monomeric subunit, here the labeling follows the usual nomenclature (b601, a602$\dots$).
  (c) Network representation of the pigments in circular layout, colors represent the typical partitioning of the pigments into communities. The widths of the links represent the strength of the couplings~$|H_{ij}|$ between nodes.
  Here the labels maintain only the ordering (b601$\to$1, a602$\to$2,$\dots$).
  (d, e, f) Partition of LHCII applying the quantum community detection algorithm in~\cite{faccin2014community}.
  Transport (for long times) and Fidelity approaches give similar results while short time transport is closer to classical community detection.
  Link width denotes the pairwise closeness of the nodes.
}
\label{fig:lhcii}
\end{figure*}

The square root of quantum Jensen-Shannon divergence has the nice property of defining a metric, allowing one to define a distance between networks. If $\rho$ and $\sigma$ are two density matrices corresponding to two networks with $N$ nodes, their Jensen-Shannon divergence is defined by
\begin{eqnarray}
\mathcal{D}_{JS}(\rho||\sigma)&=&\frac{1}{2}\mathcal{D}_{KL}(\rho||\mu) + \frac{1}{2}\mathcal{D}_{KL}(\sigma||\mu)\\
&=&S(\mu) - \frac{1}{2}[S(\rho)+S(\sigma)],
\end{eqnarray}
that is the difference between the entropy of the mixture $\mu=\frac{1}{2}(\rho+\sigma)$ and the semi-sum of the entropies of the original systems.  In the context of multilayer systems, this measure has been used to quantify the distance between layers of a multiplex network, cluster and aggregate them appropriately in order to reduce its structural complexity~\cite{dedomenico2015structural}.

These ideas have quickly found direct applications in biology. In genetic molecular systems, such as the ones described by gene-protein interactions, layers might encode different relationships among molecules---functional, e.g. additive, suppressive and other types of association, or physical, e.g. co-localization or direct interaction. The information-theoretic framework described here allowed to show that such systems exhibit a certain level of redundancy, larger than the one observed in man-made systems~\cite{dedomenico2015structural}, suggesting the existence of biological mechanisms devoted to maximize diversity of interactions.

In computational neuroscience studies, the connectome of the nematode \emph{C. elegans}---one of the most studied in the field because of its small size, with approximately 300 neuronal cells---has been mapped to a multiplex network where layers encode synaptic, gap junction, and neuromodulator interactions. Here, the analysis of reducibility revealed that the monoamine networks have a unique structure, with information complementary to that provided by neuropeptide networks~\cite{bentley2016multilayer}. The same analysis, applied to a multilayer functional representation of the human brain revealed, quantitatively, the importance of not disregarding or aggregating connectivity information for clinical classification of healthy and schizophrenic subjects~\cite{de2016mapping}.

In another application the Jensen-Shannon distance between layers of a multiplex system has been used to identify community-based associations in the human microbiome~\cite{dedomenico2016spectral}, where the microbial network corresponding to each body site is represented by a layer in a multiplex network, in perfect agreement with biological expectation~\cite{ding2014dynamics}.

The (dis)similarity between networks has also been quantified by using a combination of the classical Jensen-Shannon distance and the concept of network node dispersion, measuring the heterogeneity of a graph in terms of connectivity distance among its nodes~\cite{schieber2016quantification}. 

\heading{Degree distribution of quantum networks} Nodes in a complex network have different roles and their influence on system
dynamics can vary widely depending on their topological characteristics.
One of the simpler (and widely applied) characteristics is the degree centrality,
defined as the number of edges incident on that node.
Many real world networks have been found to follow a widely heterogeneous
distribution of degree values~\cite{albert2002statistical}.
Several models, based on mechanisms like preferential
attachment~\cite{barabasi1999emergence},
fitness~\cite{caldarelli2002fitness} or constrained random wiring~\cite{Newman03thestructure},
to mention some of them, were developed to reproduce degree distributions commonly observed in empirical systems.
Despite the complexity of the linking pattern, the degree distribution of a
network affects in a simple way the ongoing dynamics. In fact, it can be shown
that the probability of finding a memoryless random walker at a given node
of a symmetric network at the stationary state, is just proportional to the degree of such node~\cite{noh2004random}.

In~\cite{faccin2013degree} the authors consider the relationship between the stochastic and the quantum version of such processes, with the ultimate goal of shedding light on the meaning of degree centrality in the case of quantum networks.
They consider a stochastic evolution governed by the Laplacian matrix $L_{S}=\mathcal LD^{-1}$, the stochastic generator that characterizes classical random walk dynamics and leads to an occupation probability proportional to node degree.
In the quantum version, an hermitian generator is required and the authors proposed the symmetric Laplacian matrix $L_{Q}=D^{-\frac 12}\mathcal L D^{-\frac 12}$, generating a valid quantum walk that, however, does not lead to a stationary state, making difficult a direct comparison between classical and quantum versions of the dynamics.
A common and useful workaround to this issue is to average the occupation probability over time, as in Eq.~\eqref{eq:pn}. 

The generators of the two dynamics are spectrally similar (see Fig.~\ref{fig:similarity}) and share the same eigenvalues, while the eigenvectors are related by the transformation $\phi_i^C = D^{-\frac 12}\phi_i^Q$.
As a consequence, if the system is in the ground state the average probability to find the walker on a node will be the same as in the classic case, which will depend solely on the degree of each node.
For the cases in which the system is not in the ground state, it is possible to define a \emph{quantumness} measure
\begin{equation*}
  \varepsilon = 1 - \bra{\phi_0^Q} \rho_0 \ket{\phi_0^Q},
\end{equation*}
describing how far from the classical case the probability distribution of the quantum walker will be. In the case of uniformly distributed initial state $\rho_0$, this provides a measure for the heterogeneity of the degree distribution of a quantum network.

\heading{Mesoscale organization of quantum systems} Community detection, and in general mesoscopic structure detection, has been widely studied in the literature of classical complex networks~\cite{newman2012communities, fortunato2010community}.
While the definition of community ``a subset of nodes tightly connected
compared to what is expected'' is in general ill defined and part of an ongoing debate,
the number of proposed algorithms is incredibly high and still growing.

The cross-pollination of community detection with quantum mechanics is
in two levels. On the one hand, chronologically, the first attempt was to borrow tools from
quantum mechanics for applications to classical systems~\cite{horn2001clustering,weinstein2009dynamic,wittek2013quantumclustering,fanuel2016magnetic,tsomokos2011}. On the other hand an algorithm to find communities in complex quantum systems was proposed in~\cite{faccin2014community}.

In~\cite{horn2001clustering, weinstein2009dynamic}
the authors propose a method for data clustering similar to kernel density estimators, in a quantum framework.
The given data points are mapped to a Gaussian wave function and, supposing that the latter is an eigenstate for some time-independent Schr\"odingher equation:
\begin{equation*}
  H\psi = [T + V(x)]\psi = E_0\psi\,,
\end{equation*}
the minimization of the potential $V(x)$ leads to the desired clustering. An extension to dynamical quantum systems has been introduced
in~\cite{wittek2013quantumclustering}. In this case the expectation values of the
position operator evolves in time toward the closer minimum of the potential.
This formulation can leverage the
acceleration of graphics hardware. 

A method based on continuous-time quantum walks was proposed in~\cite{tsomokos2011}.
Here a node affinity measure based on the response of node
population density to link failure was given.
If the population on two nodes changes in a similar manner after link
removal, they are more likely to belong to the same community.

A magnetic Laplacian, where a magnetic field is expected to traverse all
cycles in the network, was used in~\cite{fanuel2016magnetic}.
With an approach similar to chiral walks, previously described, 
the symmetric Laplacian is amended with the original link directionality
by a phase term $e^{\pm\text{i}\theta}$, with $\theta$ being a parameter for the method, and used for community detection in directed networks, a longstanding problem in network science.

In the case of quantum systems, partitioning in modular units has been often carried out
on the basis of \emph{ad hoc} considerations.
In an effort to extend community detection to the quantum mechanics realm,  
Faccin et al.~\cite{faccin2014community} introduced several closeness matrices
inspired by different quantum quantities.
Given the Hamiltonian $H=\sum_{ij} H_{ij} \ket{i}\bra{j}$ of the quantum
system of interest, the authors consider a continuous-time random walk on the
system topology.
The first quantity is energy transport, porting to the quantum realm the concept applied in several classical algorithms where communities are interpreted as traps for the dynamical process.
In this framework, two nodes are considered to be close if, on average, their in-between transport is high.
If this average is computed over a short time period (compared to 
evolution time scales), then the closeness values are proportional to the
Hamiltonian terms $|H_{ij}|$, providing a \emph{classical} approach to
community detection (see Fig.~\ref{fig:lhcii}).
A second quantity, also proposed as a closeness measure, is related to the average
fidelity of the evolving process compared to the initial state.
In this case the localization of eigenstates is the characteristic determining
the closeness of two nodes.  These methods augment current \emph{ad hoc} approaches to partitioning nodes in quantum transport systems with enhanced methods based on community detection algorithms.

\section*{Outlook in Quantum Network Science} 

Generalization of complex network methods to the quantum setting represents a foundational advancement required to understand complexity in physical systems.  These methods represent a change of paradigm which bring several road blocks that must be faced.   Centrally, the application domain of complex network methods to quantum physics must be expanded, whereas studies in the other direction, i.e.~where methods from the quantum domain have now been ported to network science. 

From a foundational perspective, as networks necessarily represent physical systems, such systems are inherently governed by the laws of information physics.  In fact, a research line is emerging now that seeks to quantify, in terms of implicit information processing capacity, networked systems, with several applications to social, technological and biological systems~\cite{dedomenico2015structural,bentley2016multilayer,de2016mapping,dedomenico2016spectral,schieber2016quantification}.  Although this interesting direction seems promising, yet it is comparably in its infancy, whereas it is still not known how to generalize classical concepts of complexity science to the quantum domain. 

Another relevant research direction, crucial for applications in classical network science, concerns the interplay between structure and dynamics, which is almost entirely unclear in the case of quantum networks.  Although scale-free networks have been considered in the quantum setting~\cite{cuquet2009entanglement}, the result is an---albeit interesting---toy model with theoretical predictions to be verified experimentally. Therefore, further advancement along this track is of central interest, because it might play a fundamental role in quantum enhanced technology and could lead to experiments devoted to test cross-disciplinary ideas in quantum and complexity science~\cite{Perseguers2010}.  

The quest for a theoretical foundation for quantum complex networks might have a deep impact in information and communication technology. While information processing in classical systems is well controlled, it is also rather limited and quantum computing might overcome such limitations~\cite{bennett2000quantum,markov2014limits}. However, given that such systems are more sensitive to interactions with the environment, they are also more exposed to errors than their classical counterparts. Quantum error-correcting codes allow us to store and manipulate quantum information in the presence of certain types of noise that, in this context, might perturb the quantum system causing effects similar to random failures in classical complex networks. The development of quantum error correction techniques that make quantum computing and quantum communication possible can not prescind from the study of `system resilience', a topic that found uncountable applications in classical network science~\cite{albert2000error,buldyrev2010catastrophic,scheffer2012anticipating,dedomenico2014navigability,gao2016universal}. Other types of perturbations that are natural for classical systems, such as targeted attacks of network hubs~\cite{albert2000error} or cascade-based attacks~\cite{motter2004cascade}, still have no clear quantum counterpart and their study, from both theoretical and experimental perspectives, will play a key role in the development of a quantum Internet~\cite{kimble2008quantum}. In fact, it is tantalizing to think about how quantum hubs should be protected by the quantum counterpart of typical denial of service attacks. 

Continued advances in the theory of complexity in networked quantum systems will help address the challenges faced as quantum technologies scale up to commercially feasible products.  Work towards a quantum theory of complex networked systems is already opening up novel avenues when facing contemporary complexity challenges.

\vspace{0.5truecm}
\begin{small}
\noindent {\bf Author Contributions.} JDB, MF and MDD designed and wrote this review.\vspace{0.2truecm}\\
\noindent {\bf Competing Financial Interests.} The authors declare no competing financial or non-financial interests. \vspace{0.2truecm}\\
\noindent {\bf Data Availability.} No dataset were generated or analysed during the current study. \vspace{0.2truecm}\\
\noindent {\bf Acknowledgements.} JDB acknowledges the Foundational Questions Institute (FQXi, under grant FQXi-RFP3-1322) for financial support.
MF acknowledges the MOVE-IN fellowship program for financial support.
MDD acknowledges financial support from the Spanish program Juan de la Cierva (IJCI-2014-20225). The authors thank Alex Arenas and Leonie Mueck for useful feedback, and the Institute for Quantum Computing at the University of Waterloo and the Perimeter Institute for Theoretical Physics for funding and allowing us to organize the first workshop on the intersection of these topics. Diagrams are courtesy of Lusa Zheglova (illustrator).   \\ \vspace{0.2truecm} 

\end{small}


\end{document}